\documentclass[a4paper,11pt]{article}
\usepackage{pos}
\usepackage{comment}
\usepackage{tikz}
\usepackage{tikzscale}
\usepackage{neuralnetwork}

\usepackage{xcolor}
\usepackage{marginnote}

\usepackage[normalem]{ulem}



\title{Machine learning Hadron Spectral Functions in Lattice QCD\let\thefootnote\relax\footnotetext{\scriptsize This work was supported by the NSFC under the grant number 11775096, OTKA grant K123815 and the NRDI Office's MILAB Artificial Intelligence National Laboratory Program. The numerical simulations have been performed on the GPU cluster in the Nuclear Science Computing Center at Central China Normal University (NSC$^3$).}}

\author*[a]{Shi-Yang Chen}
\author[a]{Heng-Tong Ding}
\author[a,b]{Fei-Yi Liu}
\author[b]{G\'abor Papp}
\author[a]{Chun-Bin Yang}

\affiliation[a]{Key Laboratory of Quark and Lepton Physics (MOE) \& Institute of Particle Physics, \\
	Central China Normal University, Wuhan 430079, China\\}

\affiliation[b]{Institute for Physics, E{\"o}tv{\"o}s Lor\'and University, 1/A P\'azm\'any P. s\'et\'any, H-1117, Budapest, Hungary}

\emailAdd{shiyang-chen@mails.ccnu.edu.cn}
\emailAdd{fyliu@mails.ccnu.edu.cn}
\emailAdd{hengtong.ding@mail.ccnu.edu.cn}
\emailAdd{gabor.papp@ttk.elte.hu}
\emailAdd{cbyang@ccnu.edu.cn}

\abstract{
Hadron spectral functions carry all the information of hadrons and are encoded in the Euclidean two-point correlation functions. The extraction of hadron spectral functions from the correlator is a typical ill-posed inverse problem and infinite number of solutions to this problem exists. We propose a novel neural network (sVAE) based on the Variation Auto-Encoder (VAE) and Bayesian theorem. Inspired by the maximum entropy method (MEM) we construct the loss function of the neural work such that it includes a Shannon-Jaynes entropy term and a likelihood term. The sVAE is then trained to provide the most probable spectral functions.
For the training samples of spectral function we used general spectral functions produced from the Gaussian Mixture Model. After the training is done we performed the mock data tests with input spectral functions consisting 1) only a free continuum, 2) only a resonance peak, 3) a resonance peak plus a free continuum and 4) a NRQCD motivated spectral function. From the mock data test we find that the sVAE in most cases is comparable to the maximum entropy method in the quality of reconstructing spectral functions and even outperforms the MEM in the case where the spectral function has sharp peaks with insufficient number of data points in the correlator. By applying to temporal correlation functions of charmonium in the pseudoscalar channel obtained in the quenched lattice QCD at 0.75 $T_c$ on $128^3\times96$ lattices and $1.5$ $T_c$ on $128^3\times48$ lattices, we find that the resonance peak of $\eta_c$ extracted from both the sVAE and MEM has a substantial dependence on the number of points in the temporal direction ($N_\tau$) adopted in the lattice simulation and $N_\tau$ larger than 48 is needed to resolve the fate of $\eta_c$ at 1.5 $T_c$.

}

\FullConference{%
 The 38th International Symposium on Lattice Field Theory, LATTICE2021
  26th-30th July, 2021
  Zoom/Gather@Massachusetts Institute of Technology
}

\begin{document}
\maketitle

\section{Introduction}
Hadron spectral function is a quantity of great importance, since it includes all the information about hadrons, such as the dissociation temperature or transport coefficients. However, it is not directly accessible from lattice QCD computations, and is encoded in the Euclidean correlators computable on the lattice as~\cite{Ding:2015ona}
\begin{equation}
G(\tau,T)=\sum_{x,y,z} 
\left\langle J_H(0,\vec{0})J_H^+(\tau,\vec{x}) \right\rangle_T=\int_0^\infty \!\frac{\mathrm{d}\omega}{2\pi}\ K(\omega,\tau,T) \rho(\omega,T) \,,
\label{eq:Grho}
\end{equation}
where the $K(\omega,\tau,T) $ is an integral kernel expressed as,
\begin{equation}
K(\omega,\tau,T) = \frac{\cosh(\omega(\tau-\frac1{2T}))}{\sinh(\frac{\omega}{2T})}\,.
\label{eq:kernel}
\end{equation}

The extraction of the spectral functions from the correlators is a typical ill-posed inverse  problem, since typically $\mathcal{O}(10)$ data points for the correlator are available on the lattice, however, around $\mathcal{O}(1000)$ data points in the spectral function are needed in the $\omega$ space for a proper reconstruction.  Thus in principle infinite number of solutions exists. Various methods e.g., maximum entropy method (MEM)~\cite{Asakawa:2003re}, modified MEM with a different entropy term~\cite{Burnier:2013nla}, stochastic approaches~\cite{Ding:2017std}, Backus Gilbert method~\cite{Backus:1968} and Tikhonov regularization~\cite{Tikhonov:1963}, have been adopted to tackle the inverse problem.
Among these methods, the MEM, which is based on the Bayesian theorem, reconstructs the most probable spectral function and is proved to have an unique solution if it exists. 

Recently machine learning has also been applied in the reconstruction of spectral functions~\cite{Kades:2019wtd,Offler:2019eij,Zhou:2021bvw,Horak:2021syv,Shi:2021qri,Wang:2021jou}.
Variational auto-encoder (VAE)~\cite{Kingma:2019introVAE} is based on the Bayesian theorem and includes two parts: one is the encoder or recognition model and the other is the decoder or generative model. These two support each other and are connected via the distribution over the latent space $\mathcal{Z}$. The former encodes the information of the input data $x$ into a latent space $\mathcal{Z}$ and passes a conditional probability $Q(z|x)$ into a bottleneck architecture, while the latter decodes the latent space $\mathcal{Z}$ that was encoded in the bottleneck layer by the encoder to regenerate the probability $P(x|z)$ of the inputs. These results backpropagate from the neural network in the form of the loss function, during which the evidence lower bound consisting of a marginal likelihood of the input data and the closeness to the true posterior $Q_{gt}(z|x)$

In this proceedings, we present a novel neural network based on the VAE, i.e. sVAE, to reconstruct the hadron spectral function from Euclidean temporal correlation functions, and the detailed version of the current proceedings can be seen in~\cite{Chen:2021giw}. Inspired by the MEM, the loss function in the sVAE is constructed such that it includes a Shannon-Jaynes entropy term and a likelihood function. The `s' in the name of `sVAE' is then attributed to the entropy term in the loss function. The sVAE is trained such that it learns to obtain the most probable spectral function by balancing the closeness to prior information of spectral function encoded in the entropy and the correlator data in the likelihood.
In section~\ref{sec:sVAEintro} we present the details of the loss function in the sVAE, the topology of the neural network and the training samples of spectral functions. In section~\ref{sec:mocktests} we show the mock data tests of the sVAE. In section~\ref{sec:lqcdresults} we show the application of the sVAE to the correlators computed in the quenched lattice QCD. We finally summarize our results in section~\ref{sec:summary}.

\section{sVAE}
\label{sec:sVAEintro}
In this section, we present the details on the sVAE, including its loss function, topology and the used training samples of spectral functions. To avoid the clutter we suppress the arguments $\tau$ and $T$ of the temporal correlator $G$, and $\omega$ and $T$ of the spectral function $\rho$.

\subsection{The loss function and topology of the sVAE}
\begin{figure}[!thpb]
	\centering
	\includegraphics[]{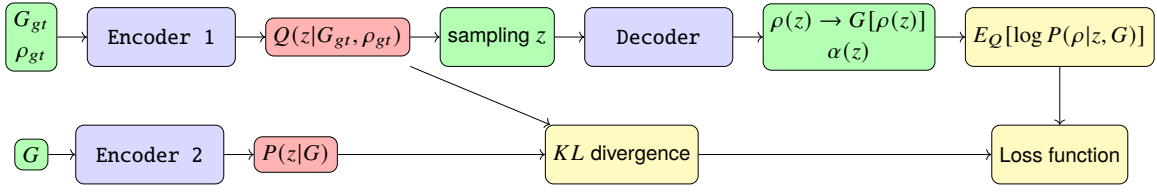}
	\caption{Training process involving the encoders and decoder~\cite{Chen:2021giw}.
	}
	\label{fig:section3fig1}
\end{figure}

The loss function of the sVAE is expressed as follows~\cite{Chen:2021giw},
	\begin{equation}
		\mathfrak{L} = - E_{Q(z|\rho_{gt},G_{gt})}\bigg[ \log \frac{P(\rho|z)P(G|\rho, z)}{P(G|z)}\bigg] + KL\bigg(Q(z|\rho_{gt},G_{gt})\| P(z|G) \bigg),	
		\label{eq:loss}
	\end{equation}
where $Q(z|\rho_{gt},G_{gt})$ is the distribution of latent code $z$ given $\rho_{gt}$ and $G_{gt}$ as the ``ground-truth'' values of spectral and correlation function, respectively. $P(z|G)$ gives the probability of $z$ given the correlator $G$. These two probabilities are estimated by two networks of ``Encoder 1'' and ``Encoder 2'' (see Fig.~\ref{fig:section3fig1}), respectively. ``Encoder 1'' encodes the $(\rho_{gt}, G_{gt})$ pair into latent space $\mathcal{Z}$, while ``Encoder 2'' maps the stochastic correlation functions to the same latent space $\mathcal{Z}$
\begin{equation}
KL\bigg(Q(z|G_{gt},\rho_{gt})\| P(z|G)\bigg) = \int dz~ Q(z|G_{gt},\rho_{gt})\log\bigg( \frac{Q(z|G_{gt},\rho_{gt})}{P(z|G)} \bigg).
\label{eq:KL-term}
\end{equation}

The prior probability $P(\rho|z)$ in Eq.~\eqref{eq:loss} is parameterized in terms of the Shannon-Jaynes entropy $S$ as~\cite{JARRELL1996133,Asakawa:2000tr},
\begin{align}
P(\rho|z) = \frac{1}{Z_S}e^{S},
\label{eq:prior}
\qquad \mbox{where} \quad
S = \sum\limits_l^{N_\omega} \left(\rho_l(z) - \rho_{gt,l} - \rho_l(z)\log\left(\frac{\rho_l(z)}{\rho_{gt,l}}\right) \right)\,.
\end{align}
Here $Z_S\approx (2\pi)^{\frac{N_\omega}{2}}$ is normalization factor, and $N_{\omega}$ is the number of data points used to sample the frequency space $\omega$. $P(G|\rho, z)$ gives the probability to reconstruct the measured correlator $G$ parameterized in terms of the likelihood function~\cite{JARRELL1996133,Asakawa:2000tr},
\begin{equation}
P(G|z, \rho)  = \frac{1}{Z_L}e^{- L} \,
\label{tab:section2eq4}
\qquad \mbox{with}\quad
L =\sum\limits_{j=\tau_{min}}^{N_{\tau}/2} L_j =\sum\limits_{j=\tau_{min}}^{N_{\tau}/2} \frac{\Big(\hat{G}_j\big[\rho(z)\big]- G_j\Big)^2}{2\alpha_j^2(z)G_j^2}
\end{equation}
having a form of a likelihood function, and $Z_L = \prod_j^{N_{\tau}} \sqrt{2\pi\alpha_j^2(z)G_j^2}$.
Here $\hat{G}\big[\rho(z)\big]$ is the correlator reconstructed from $\rho(z)$ according to the relation shown in Eq.~\ref{eq:Grho}, and $\alpha(z)$ is a weight parameter to adjust the weight between the entropy term and likelihood term. Through the ``Encoder 1" and ``Encoder 2" we obtain the probability of $z$, i.e. $Q(z|G_{gt},\rho_{gt})$ and $P(z|G)$, respectively, we then construct a ``Decoder'' network (cf. Fig.~\ref{fig:section3fig1}). Through the ``Decoder" the spectral function $\rho(z)$ and the corresponding weight $\alpha(z)$ are obtained for the evaluation of the loss function. 

\begin{figure}[!thpb]
	\centering
	\includegraphics[]{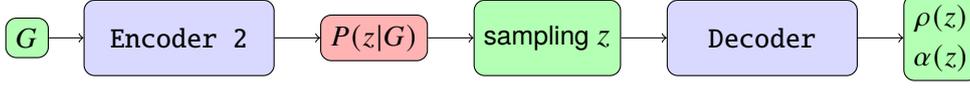}
	\caption{Reconstruction process~\cite{Chen:2021giw}.}
	\label{fig:section3fig2}
\end{figure}

During the training the first term in Eq.~\eqref{eq:loss} aims to minimize $S-L$ by balancing the closeness of $\rho$ to the ground truth value of spectral function $\rho_{gt}$ and to correlator data $G$. This resembles the MEM. While the second term, the KL term, enforces the $z$ distribution obtained from ``Encoder 2'' to approach the one from  ``Encoder 1''.
After the training, we drop ``Encoder1'' , and only the ``Encoder 2''-``Decoder'' chain is used for the reconstruction as shown in Fig.~\ref{fig:section3fig2}.
Since ''Encoder 2'' results in a probability distribution of the latent variable $z$ given the stochastic correlator $G$, the reconstructed spectral function can be obtained with $N_s$ samples as follows~\cite{Chen:2021giw},

\begin{equation}
\begin{aligned}
\tilde{\rho}&=  \int\!\! dz\ \rho(z) P(G|\rho, z)P(z|G)  
                    \to \sum\limits_{n=1}^{N_s} \frac{\rho(z) P(G|\rho, z_n)P(z_n|G)}{\sum\limits_{k=1}^{N_s} P(G|\rho, z_k)P(z_k|G)} \,.
\end{aligned}
\label{eq:rho}
\end{equation}

\subsection{Training samples}
\begin{figure}[!htbp]
\centering
	\includegraphics[width=0.45\textwidth]{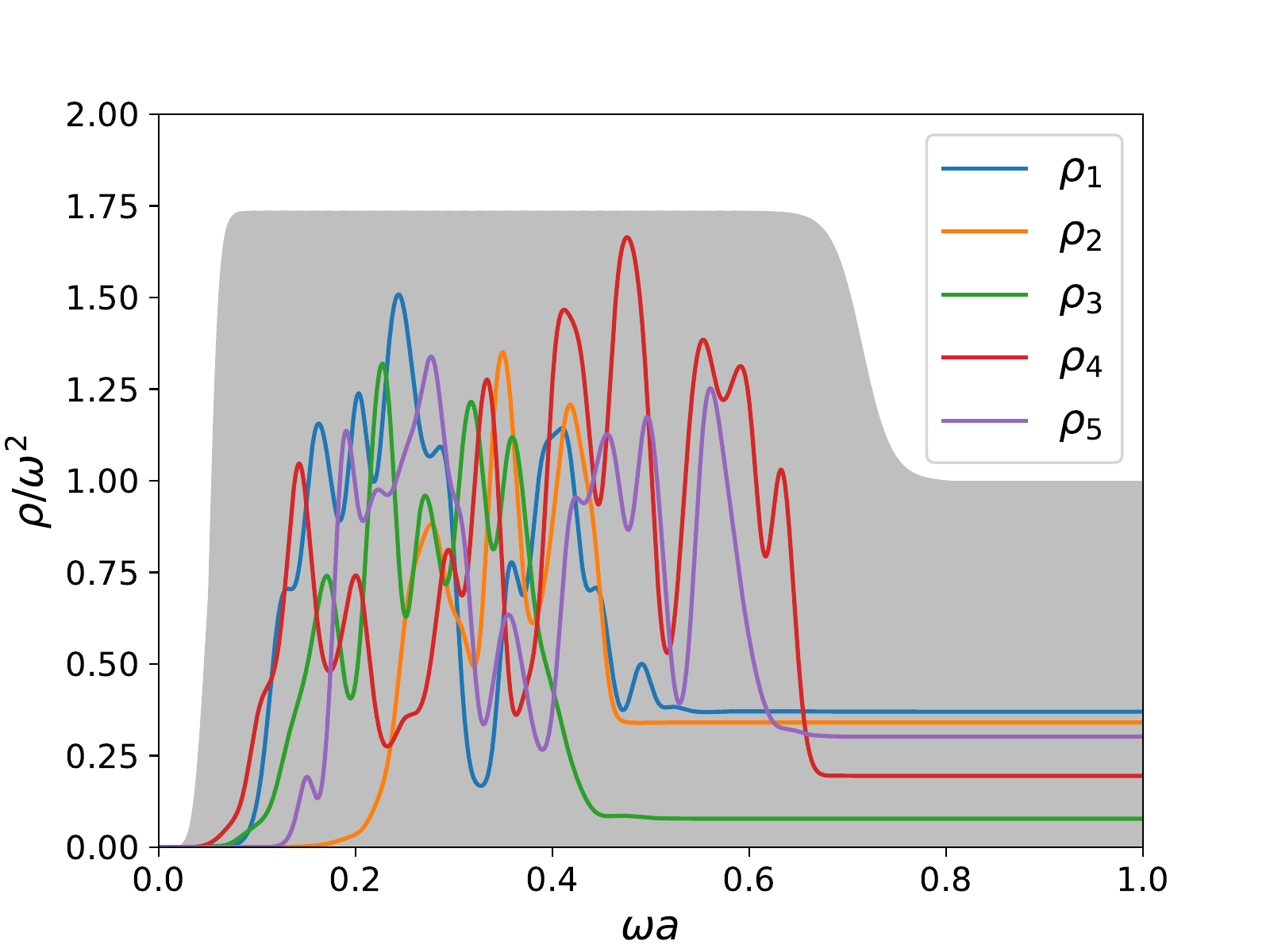}
	\caption{5 samples of the training spectral function $\rho$. The grey shadow area shows the full sampled range~\cite{Chen:2021giw}.
	}
	\label{fig:training_sample}
\end{figure}

The training spectral function is constructed using a Guassian mixture model. In practice the spectral function consists of a smoothed sum of many Gaussian peaks with randomly chosen width and height parameter, and also a continuum (constant) part~\cite{Chen:2021giw}. The peak locations of Gaussian peaks  are chosen uniformly within [0,4] in the $\omega$ space with number of points $N_\omega=10000$. Five sampled training spectral functions and the covered region are illustrated in Fig.~\ref{fig:training_sample}.

The training sample $\rho_{gt}$ and resulting temporal correlator $G_{gt}(\tau)$ are fed to ``Encoder 1" (cf. Fig.~\ref{fig:section3fig1}). Only $\tau \leq N_\tau/2$ is used due to the symmetry (Eq.~\eqref{eq:kernel}), and $\tau \ge 4$ is set due to lattice cutoff effects.
For each $G_{gt}(\tau)$, a Gaussian noise is added to construct the stochastic data $G(\tau)$ so as to mimic the error from computations in lattice QCD. Thus $G$ is sampled from a Gaussian distribution with its mean value $G_{gt}$ and its variance having the following form,
\begin{equation}
\sigma_{lat}(\tau)/G_{lat}(\tau)\times {G}_{gt}(\tau),
\label{eq:noise_level}
\end{equation}
where $\sigma_{lat}(\tau)/G_{lat}(\tau)$ is the lattice noise ratio level at $\tau$ (typically $1.5\%$ at $\tau=N_\tau/2$). The resulting stochastic data $G(\tau)$ is fed to ``Encoder 2" (cf. Fig.~\ref{fig:section3fig1}).

\section{Mock data tests}
\label{sec:mocktests}
\begin{figure}[!htp]
	\centering
	\includegraphics[width=0.45\textwidth]{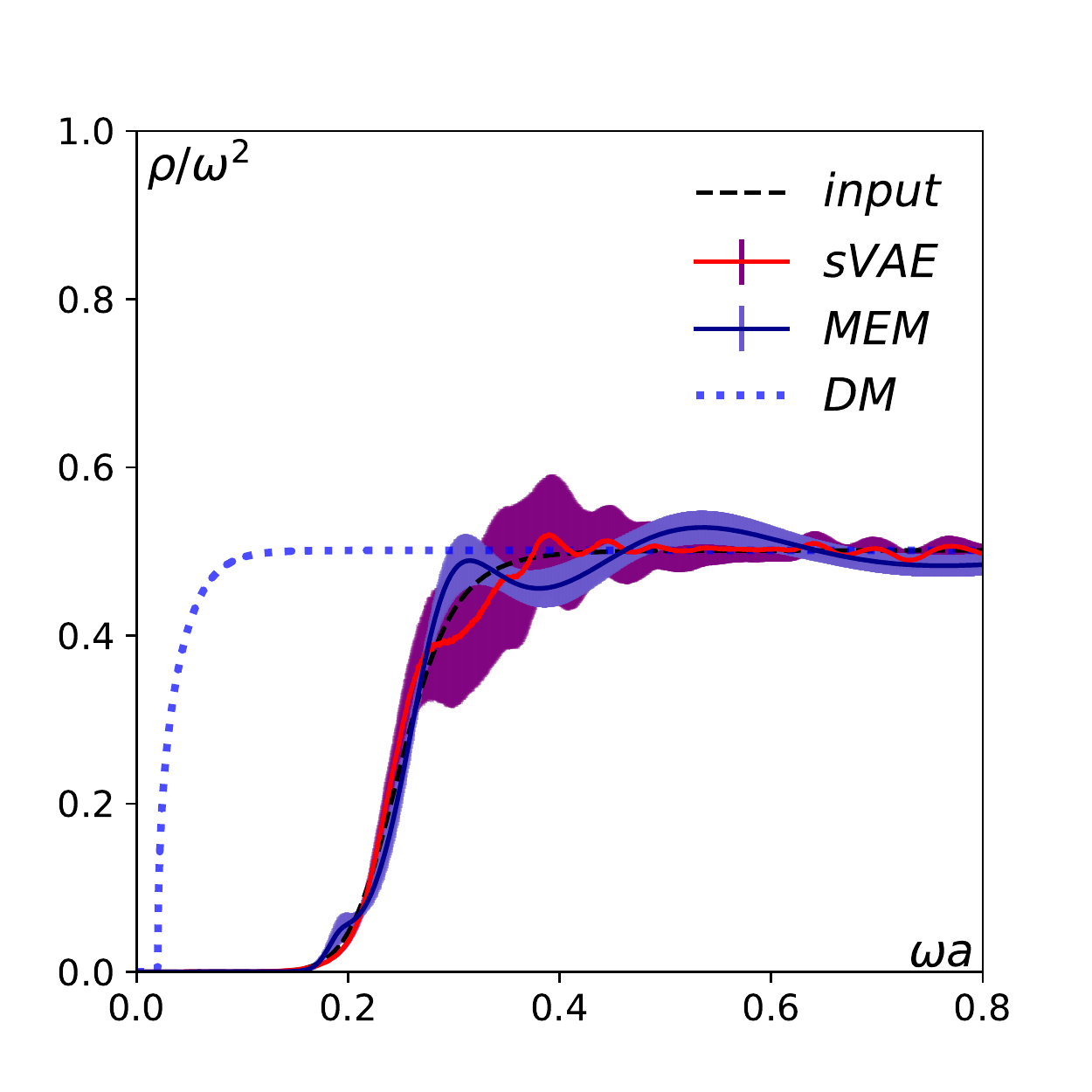}
	\includegraphics[width=0.45\textwidth]{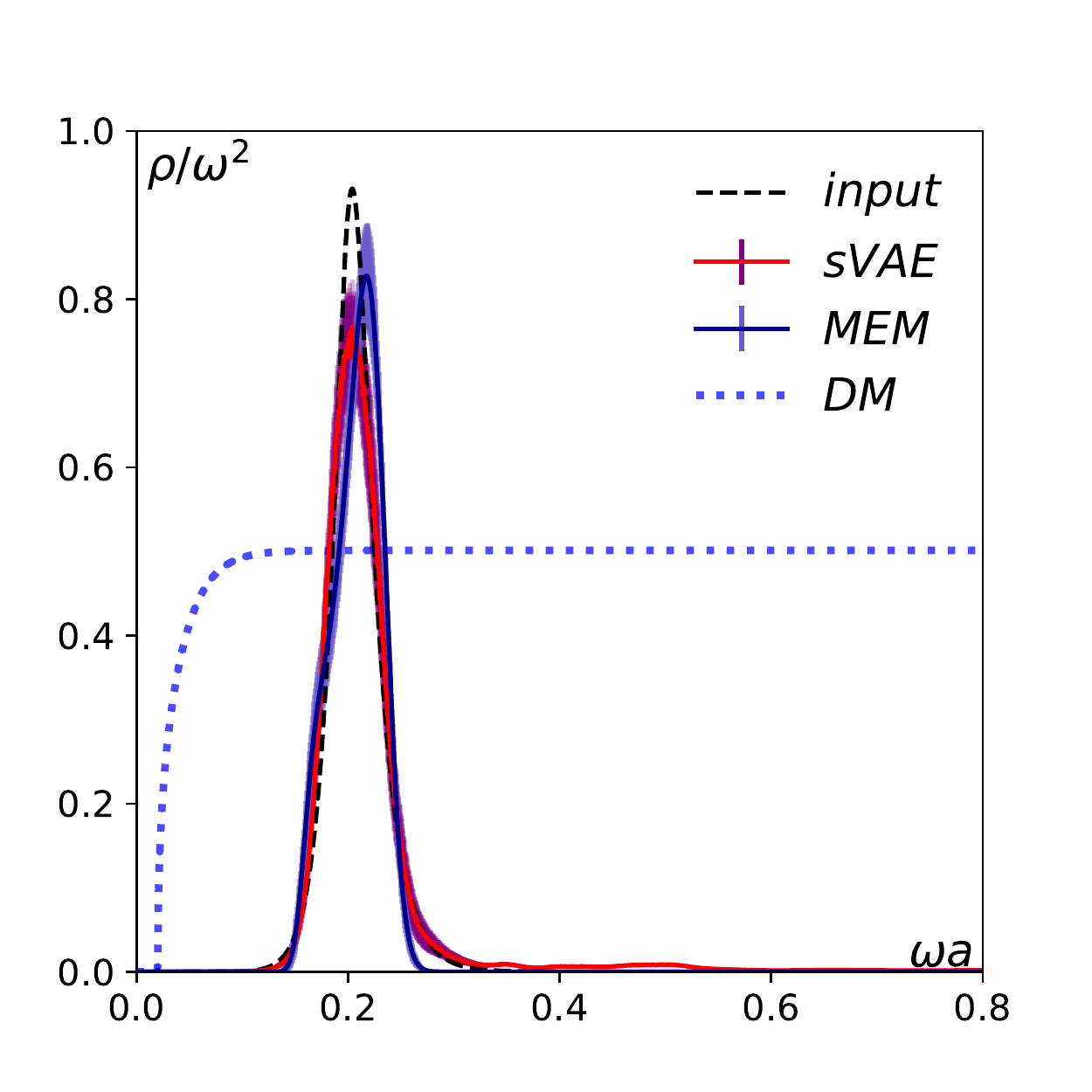}
	\caption{Mock data tests with an input spectral function containing only a continuum part (left) and with only a resonance peak (right). The input mock spectral function is denoted by the black dashed line, and the mean value and uncertainty of the output spectral function from the trained sVAE are represented by the red solid line and purple band, respectively. The blue line and blue band are the mean value and uncertainty of the output spectral function from MEM, respectively. The corresponding blue dotted line is the default model of MEM. The results are obtained with $N_\tau=96$ and are taken from Ref.~\cite{Chen:2021giw}. }
	\label{fig:case12_Nt96}
\end{figure}

After the training with the training samples of spectral functions illustrated in Fig.~\ref{fig:training_sample} is done, we first make two mock data tests with only a free continuum spectral function and only a resonance peak in the input spectral function. The results from both the sVAE and MEM in these two cases are shown in the left and right plots of Fig.~\ref{fig:case12_Nt96}, respectively. It can be seen that the results from the sVAE and MEM in these two cases are comparable, and sVAE seems outperform MEM in the reconstruction of the peak location of the resonance peak shown in the right plot of Fig.~\ref{fig:case12_Nt96}.
	
\begin{figure}[!htp]
	\centering
	\includegraphics[width=.9\textwidth]{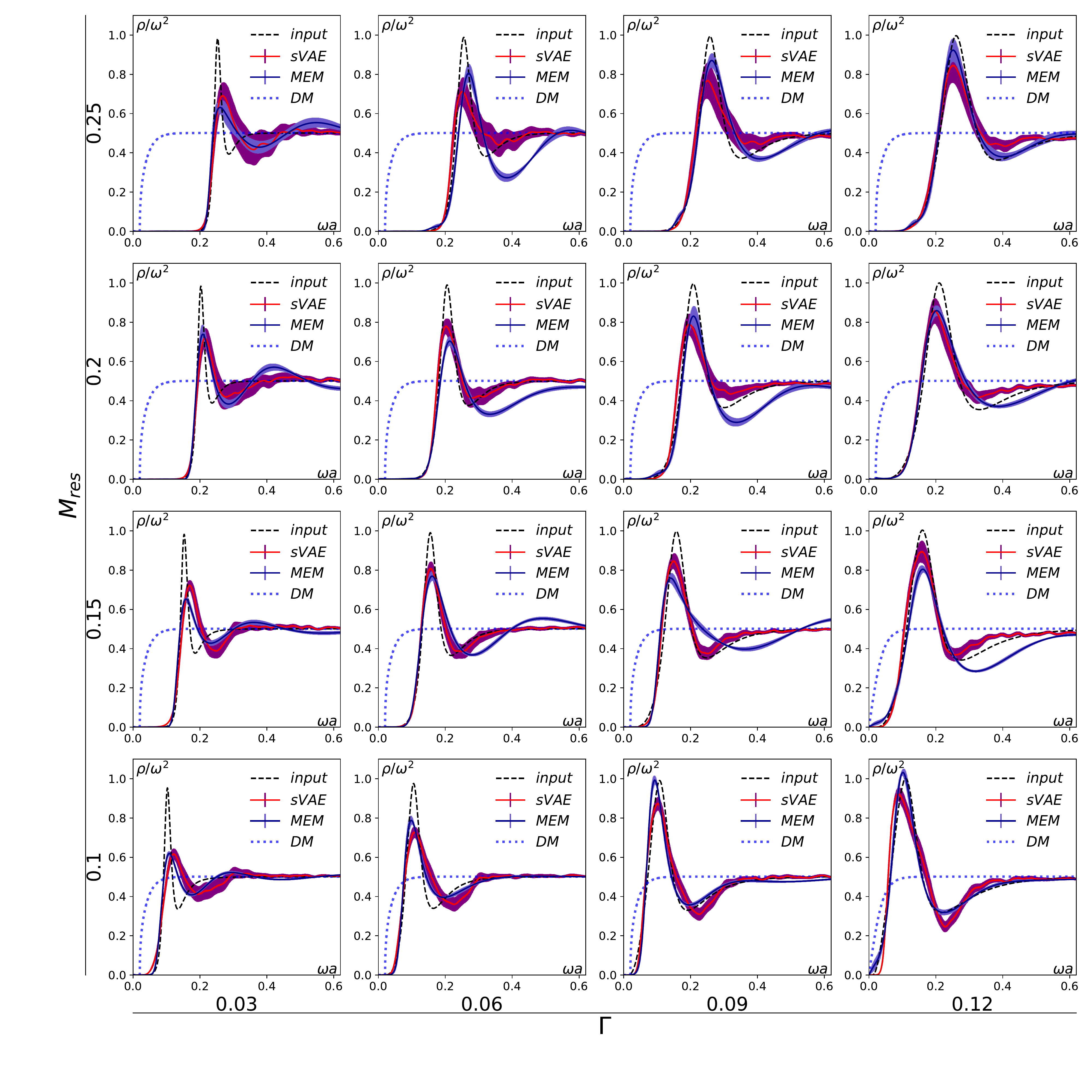}
	\caption{Mock data tests with each input spectral function containing a resonance peak and a continuum part with $N_{\tau}=96$~\cite{Chen:2021giw}. The legends are the same as those in Fig.~\ref{fig:case12_Nt96}.}
	\label{fig:case3_Nt96}
\end{figure}

We make further mock data tests with a more complicated input spectral function. The input spectral function now consists of a continuum part and one resonance (Breit-Wigner) peak. In this test we study the dependence of the reconstruction quality on the peak location and width of the resonance peak in the input spectral function. In Fig.~\ref{fig:case3_Nt96} we increase the peak width of the input spectral function ($\Gamma$) from left to right horizontally, and increase the peak location ($M_{res}$) from bottom to top vertically. As seen from Fig.~\ref{fig:case3_Nt96} both the sVAE and MEM cannot reconstruct the spectral function in a satisfactory way when the peak width is small, e.g. $\Gamma=0.03$. With $\Gamma\geq 0.06$ both the sVAE and MEM can reconstruct the spectral function reasonably well and yield comparable results.

\begin{figure}[!htp]
	\centering
	\includegraphics[width=0.45\textwidth]{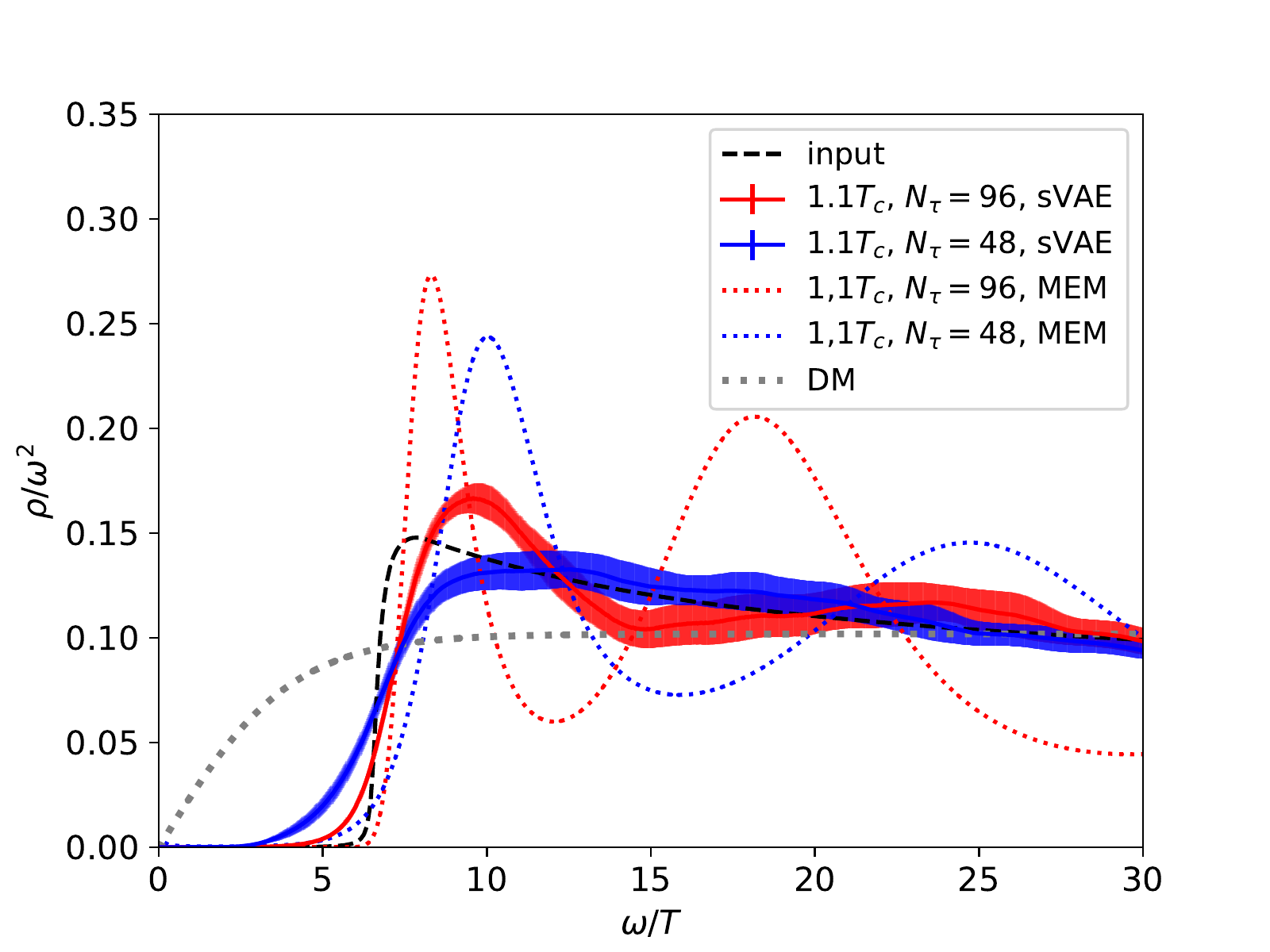}
	\includegraphics[width=0.45\textwidth]{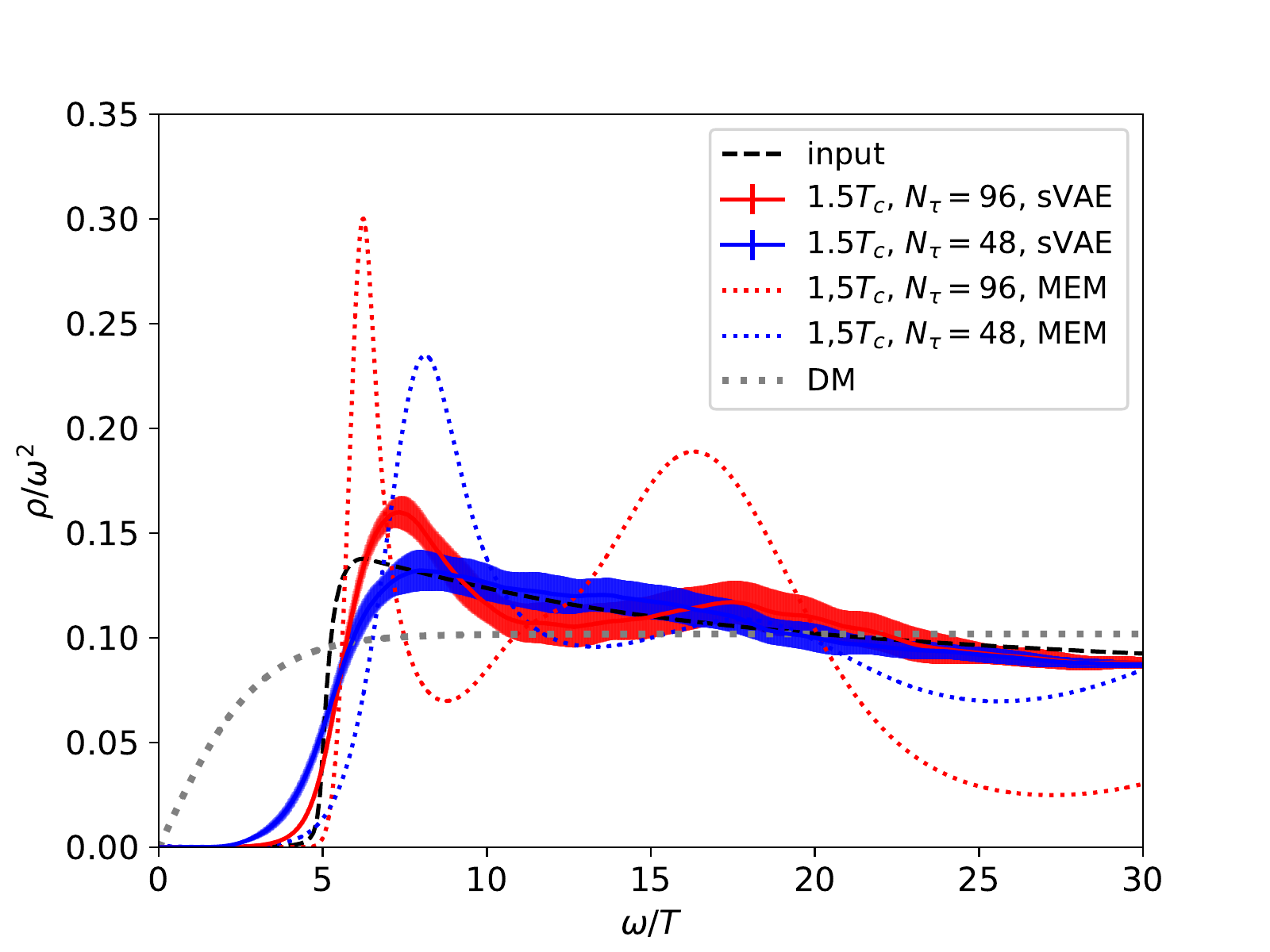}
	\caption{Mock data tests with input spectral functions obtained from 2-loop perturbative NRQCD at T=1.1$T_c$ (left) and 1.5$T_c$ (right)~\cite{Chen:2021giw}. In both plots the input spectral function is denoted by the black dashed line, and the red solid line and blue solid line with surrounding bands denote the results obtained from the sVAE for $N_\tau=96$ and $N_\tau=48$, respectively. The dotted lines with the same color denote results obtained from the MEM with the default model denoted as grey dotted lines.}
\label{fig:case4NRQCD}
\end{figure}

As understood from the tests shown in Fig.~\ref{fig:case3_Nt96} the reconstruction quality crucially depends on how rapidly the peak rises or how small the peak width is, we make a further mock data test with a NRQCD motiviated spectral function as the input spectral function. The NRQCD spectral function increases rapidly around the threshold ($\omega/T \sim 6$) and then slowly decreases. We show the reconstructed spectral functions obtained from both the sVAE and MEM in Fig~\ref{fig:case4NRQCD}. It turns out that both the sVAE and MEM cannot reproduce the input spectral function, and the MEM even produce fake peaks which do not exist in the input spectral functions.

\section{Analysis on lattice QCD data}
\label{sec:lqcdresults}
In this section, we apply the sVAE to reconstructing spectral functions from Euclidean two-point correlation functions computed in quenched lattice QCD. We focus on the charmonium correlator in the pseudo-scalar channel. The correlator data analyzed here comes from~\cite{Ding:2012sp}, which were computed on $128^3\times 96$ and $128^3\times 48$ in quenched lattice QCD corresponding to temperatures 0.75 $T_c$ and 1.5 $T_c$ with inverse lattice spacing $a^{-1}=18.97$ GeV, respectively. To avoid lattice cut-off effects $\tau_{min}$ is set to 4, and the frequency space has been cut at $\omega_{max} a=4$ due to the lattice cutoff in the free spectral function~\cite{Karsch:2003wy,Aarts:2005hg}. 
 \begin{figure}[!htp]
	\centering
	\includegraphics[width=0.32\textwidth]{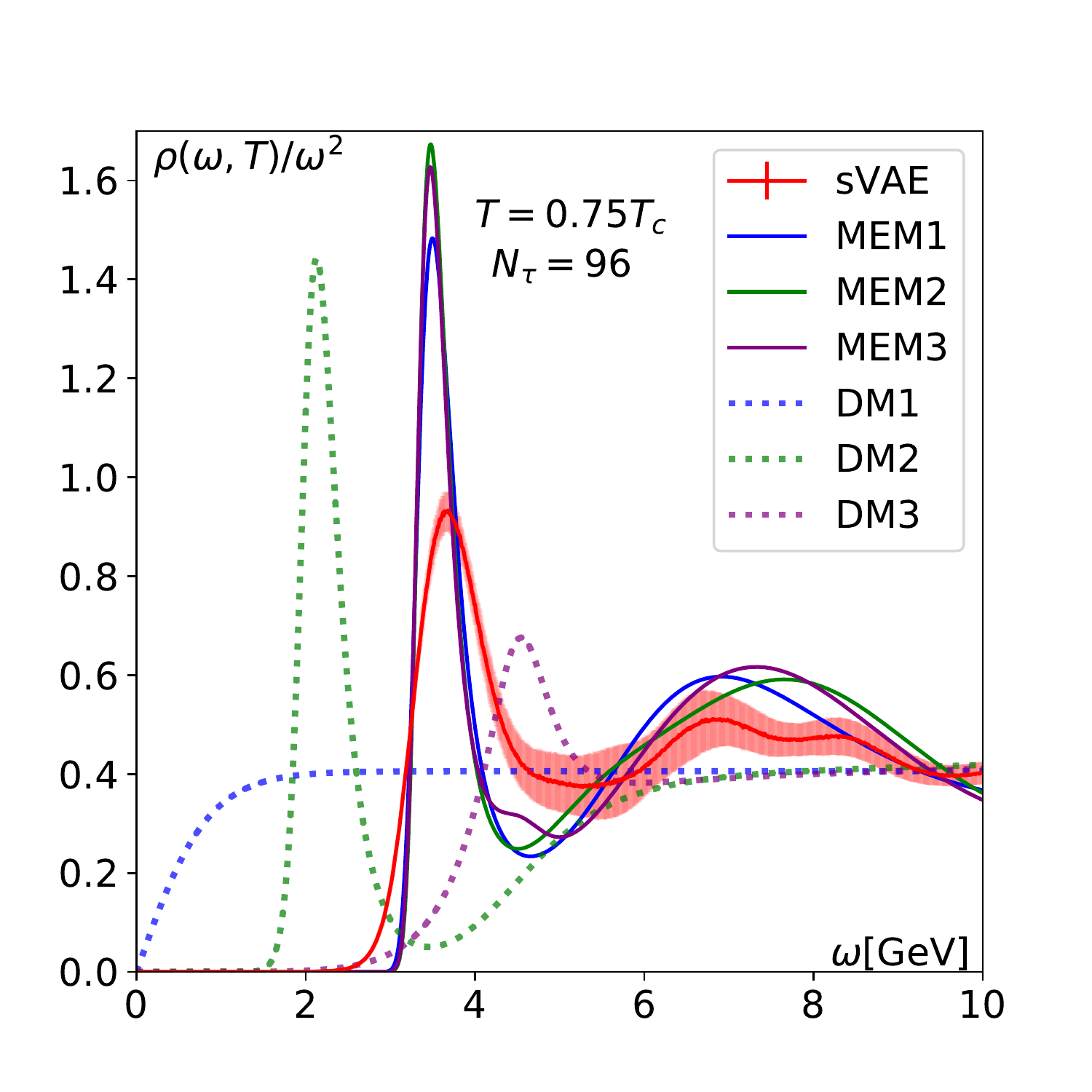}
	\includegraphics[width=0.32\textwidth]{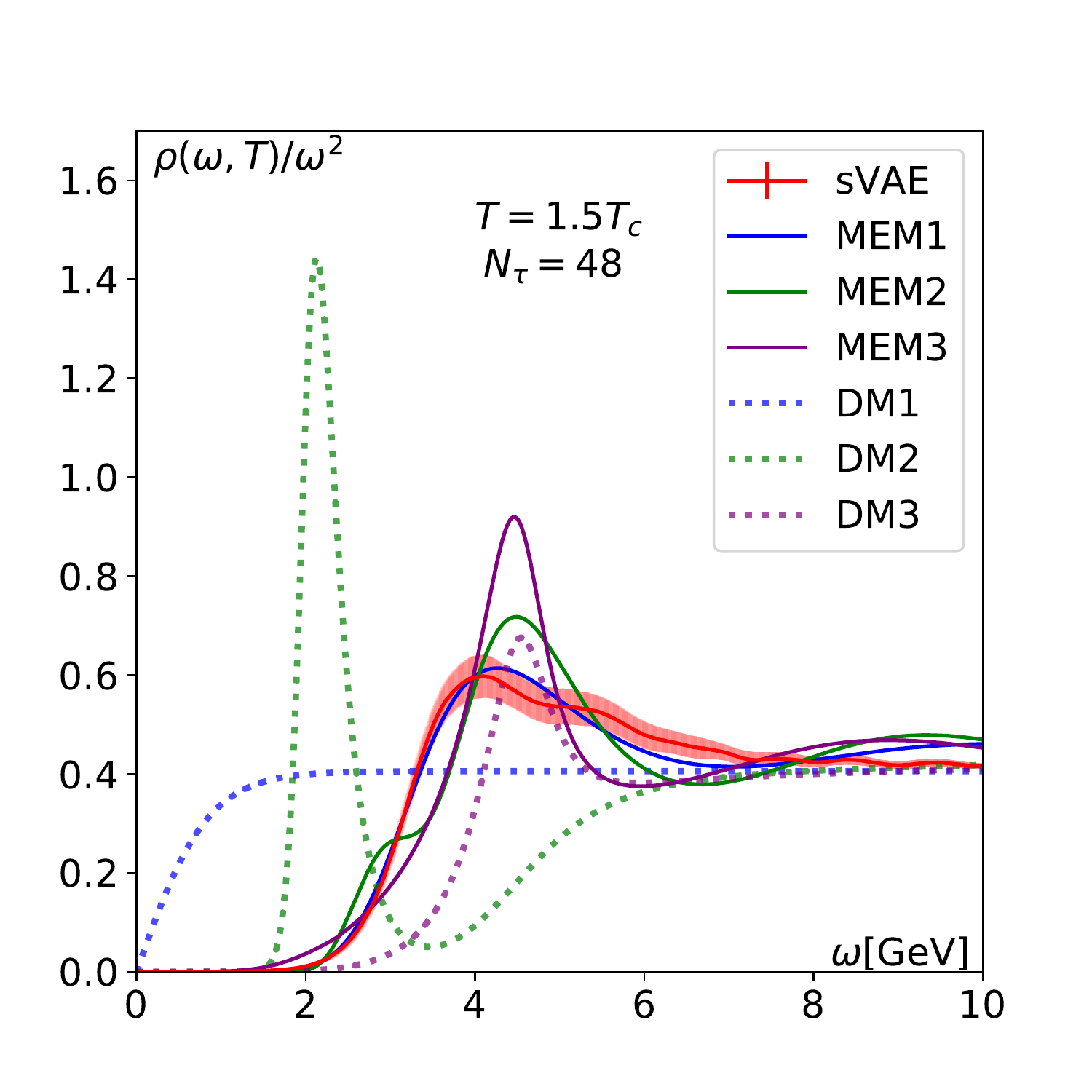}
	\includegraphics[width=0.32\textwidth]{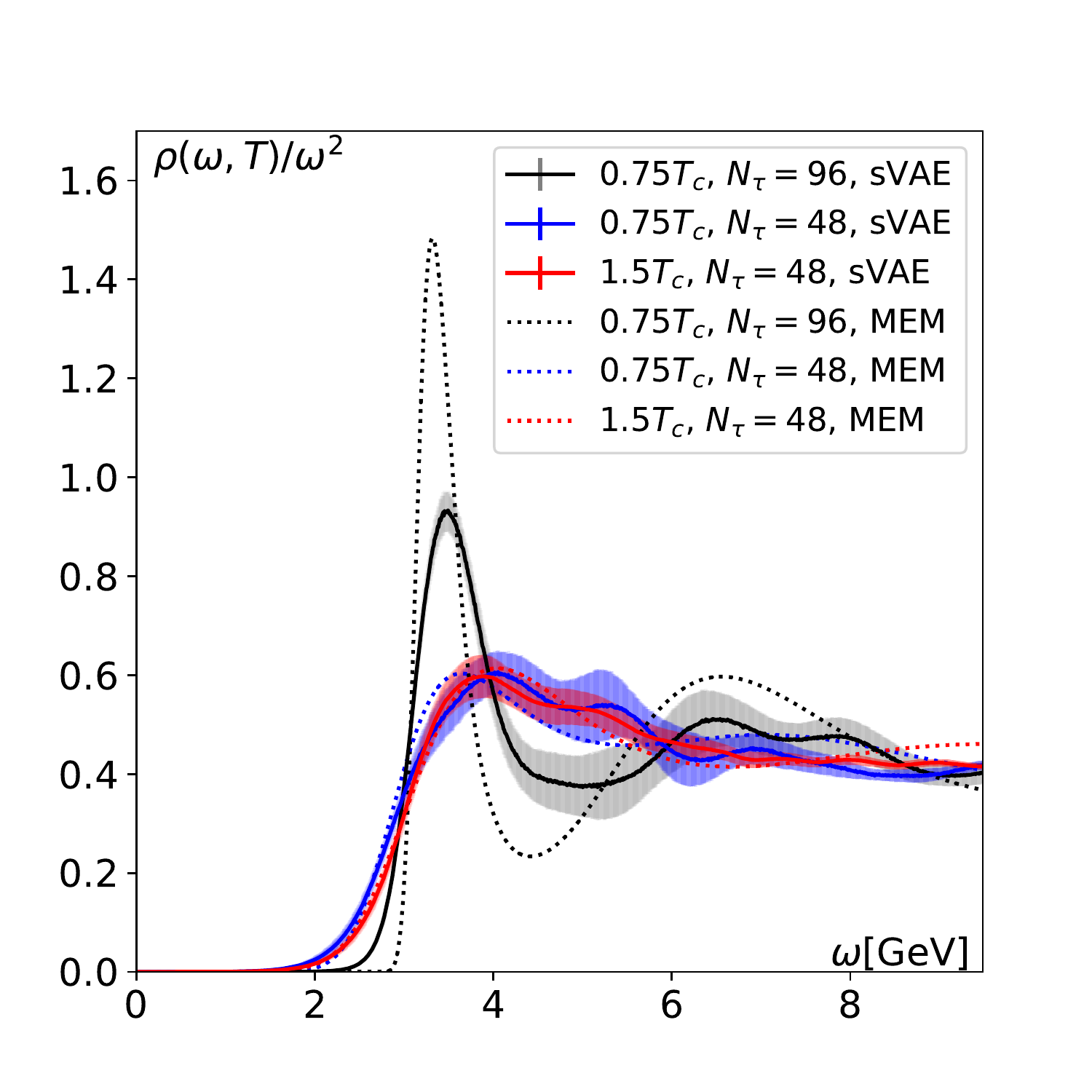}
	\caption{Left: Spectral functions at 0.75 $T_c$ obtained from the sVAE and MEM with $N_\tau=96$.
	Middle: Same as the left plot but obtained at 1.5 $T_c$ with $N_\tau=48$. Right: A combined plot of the left and right ones. Here the MEM results obtained only using `DM1' are shown. In addition spectral functions obtained from the sVAE and MEM at 0.75 $T_c$ with $N_\tau=48$ are also shown.
	}
\label{fig:lattice_spf}
\end{figure}

In the left plot of Fig.~\ref{fig:lattice_spf} we show the reconstructed spectral functions from the sVAE at 0.75 $T_c$. For comparison results obtained using the MEM are also shown with three different default models (DM). It can be observed that the results from the MEM are almost independent of DMs and the reconstructed peak location lies at $\omega\approx 3.34$ GeV. The mass of $\eta_c$ as read-off from this peak location is consistent with the pole mass extracted from the spatial correlation function at 0.75 $T_c$~\cite{Ding:2012sp}. On the other hand, the peak location obtained from the sVAE is about 5\% larger than that from the MEM. And the peak height obtained from the sVAE is always smaller than that from the MEM. Despite these differences, the resonance peak structure obtained from the sVAE and MEM is comparable.

 
 At 1.5 $T_c$ the reconstruction becomes harder since the number of data points become half of that at 0.75$T_c$. As seen from the middle plot of Fig.~\ref{fig:lattice_spf} the spectral function obtained from the sVAE does not possess a remark peak structure and is only comparable with the MEM result obtained using a featureless DM (`DM1'). For the MEM results obtained using `DM 2' and `DM 3' the peak location shifts to a much larger $\omega\approx 4.26$ GeV.
 
To make direct comparison between the results shown in the left and right plots of Fig.~\ref{fig:lattice_spf} we summarize them in the right plot of Fig.~\ref{fig:lattice_spf} with MEM results only from `DM1'. As mentioned in the previous paragraph it seems that significant thermal effects are attributed to the change in the spectral function at 1.5 $T_c$ from 0.75 $T_c$. However, we also need to check the $N_\tau$ dependence in the reconstruction of the spectral functions from both methods. We thus looked at the reconstructed correlator at 0.75 $T_c$. This reconstructed correlator encodes the same spectral function at 0.75 $T_c$ but has a smaller number of points in the temporal direction, i.e. $N_\tau=48$. It can be clearly seen that there exits a large $N_\tau$ dependence in both sVAE and MEM during the extraction of $\rho(\omega,0.75T_c)$. It is thus hard to tell the fate of $\eta_c$ at 1.5 $T_c$ based on the current lattice QCD data.
 
  

\section{Summary and conclusion}
\label{sec:summary}
In this proccedings, we presented a novel neural network (sVAE) based on the the variational auto-encoder and Bayesian theorem. The sVAE is trained such that it provides the most probable spectral function by balancing the prior and the likelihood in the loss function. For the training samples we used general spectral functions constructed using the Gaussian mixture model.
In the mock data tests, we found that results from the sVAE are comparable to those from the MEM, and in some cases sVAE even outperforms MEM. In the application to the lattice QCD correlator data, the peak locations of the results at 0.75 $T_c$ from sVAE and MEMs are consistent with each other, and the peak height in the output spectral function from the sVAE is always smaller than that from the MEM. At 1.5 $T_c$, the $N_{\tau}$ dependence of spectral functions still imposes a challenge to extract a reliable spectral function. Thus, to determine the fate of $\eta_c$ at 1.5 $T_c$, $N_{\tau}$ larger than 48 is probably needed.



\bibliographystyle{ieeetr} 
\bibliography{mlspf}

\end{document}